\documentclass[11pt,twoside]{article}
\usepackage{asp2010}
\usepackage{epsf}
\usepackage{lscape}
\usepackage{natbib}
\usepackage{graphicx}

\resetcounters

\begin{document}

\title{The injection and feedback of Cosmic Rays in large-scale structures}
\author{F. Vazza$^{1,3}$, M. Br\"{u}ggen $^1$, C. Gheller $^2$ and G. Brunetti $^3$
\affil{$^1$ Jacobs University Bremen, Campus Ring 1, 28759 Bremen, Germany\\
$^2$ Swiss National Supercomputing Center, Via Cantonale, CH-6928 Lugano, Switzerland\\
$^3$ INAF, Istituto di Radio Astronomia, 41029 Bologna, Italy}}

\begin{abstract}
We present the numerical implementation
of run-time injection of Cosmic Rays energy, their spatial
advection and their dynamical feedack on baryonic gas in the cosmological grid
code ENZO. 
We discuss the results of its application to large-scale simulations showing that the CR energy inside clusters of galaxies is small compared to
the gas energy (less than a few percent), while the ratio
is larger near the accretion regions of clusters and filaments ($\sim 0.1-0.3$). CR feedback has a small, but significant
impact on the X-ray emission and Sunyaev-Zeldovich effect from clusters.
\end{abstract}

\section{Introduction}
The detection of large-scale non-thermal emissions in galaxy clusters proves the existence
of magnetic fields and relativistic particles (CR) in the intra cluster medium
(e.g. \citealt{fe08}). Merger events in clusters drive shock waves in the intra cluster medium, and this can inject primary CR particles
and may offer  a simple explanation of the observed connection between large-scale radio sources with arc-like shapes (e.g. "radio relics", \citealt{br11} for a recent review) and the underlying disturbed cluster morphology. The mechanism to convert kinetic energy from the $\sim$ Mpc scale of shock waves into the acceleration of CRs may be explained by the Diffusive Shock Acceleration (DSA, e.g. \citealt{bell78}; \citealt{kj07}).
According to this model, a pool of supra-thermal particles
is produced by  collision-less shocks in rarefied space plasmas, and are accelerated to higher 
energies by the interactions with Alfv\'{e}n waves in the post-shock region.
We present our implementation of a model of DSA in the public release of the cosmic code ENZO 1.5 (see Sect.\ref{sec:methods}), simulating the impact of CR physics in the evolution of large-scale structures (see Sect.\ref{sec:results}).

\begin{figure}[!ht]
\includegraphics[width=0.9\textwidth]{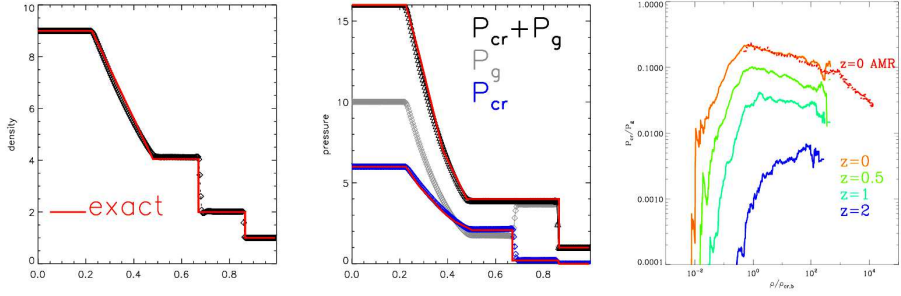}
\caption{Left and central panel: 1--D shock tube test for an initial mixture of gas and CRs (shown are density, gas, CRs and total pressure). The red lines show the exact solution from  \citet{mi07}. Right panel: average pressure ratio $P_{\rm cr}/P_{\rm g}$ as a function of gas overdensity for a cosmological run at fixed grid resolution ($\Delta x=156$ kpc/h) for different cosmic epochs. Additionally shown is dotted-red is the result at z=0 for an AMR resimulation of a cluster, with a maximum resolution of $\Delta x=25$ kpc/h.}
\label{fig:fig1}
\end{figure}

\section{Numerical methods}
\label{sec:methods}
All simulations presented in this work were performed using the 
grid-based and adaptive mesh refinement code ENZO \citep{no07,co11}. 
ENZO is currently developed developed by the Laboratory for Computational  Astrophysics at the University of California in San Diego (http://lca.ucsd.edu).
In our work, we implemented the treatment of CR energy injected at shocks 
by adding several routines, physical fields and new sets of equations
to the public 1.5 release of ENZO, following a two-fluid model approach (e.g. \citealt{kj07}).

An on-the-fly shock finder was developed to detect shocks in
simulations at each time step. This shock finder is  based on a 3-D analysis of gas pressure, similar to other methods in the literature(e.g. \citealt{ry03}; \citealt{sk08}; \citealt{vbg09} ). Once the Mach number, $M$, is measured, we estimate the total energy flux of accelerated CR protons
as: $\phi_{\rm cr} = \eta (M) \cdot \rho_{\rm u} v_{\rm s}^{3}/2$ ($\rho_{\rm u}$ is the pre-shock gas density,  $v_{\rm s}$ is the shock velocity, and the function $\eta(M)$ models the efficiency of injection, as in \citet{kj07}). We inject  CR energy in each post-shock cell as $E_{\rm cr}=\phi_{\rm cr} \Delta t \Delta x^{2}$, where $\Delta t$ is the time step and $\Delta x$ is the cell size; a corresponding amount of energy is also decreased in the post-shock region to conserve total energy.
The CR energy is advected with the flow assuming no diffusion and no energy losses for CRs, using an adiabatic index of $\gamma_{\rm cr}=4/3$. 
The total pressure $P=P_{\rm g}+P_{\rm cr}$  (where $P_{\rm cr}=(\gamma_{\rm cr}-1) E_{\rm cr}$ and $P_{\rm g}$ is the gas pressure) is fed into the PPM Riemann solver of ENZO (rather than the gas pressure). The composite fluid gas+CR in the simulation obeys an effective adiabatic index, $\gamma_{\rm eff} \leq \gamma=5/3$. 
To preserve the stability of the method, we also enforce an additional condition on the time step (which avoids too strong a modification of the post-shock during a single time step). For a full description of the adopted methods, and of their tests against analytical solutions we refer the reader to a forthcoming paper (Vazza et al., in prep). In Fig.\ref{fig:fig1},  we show the comparison of a 1--D shock tube test and the numerical solution obtained with the Glimm-Godunov's method by \citet{mi07}, showing the very good performance of our new scheme.

\section{Results}
\label{sec:results}

\begin{figure}[!ht]
\includegraphics[width=0.95\textwidth,height=0.9\textwidth]{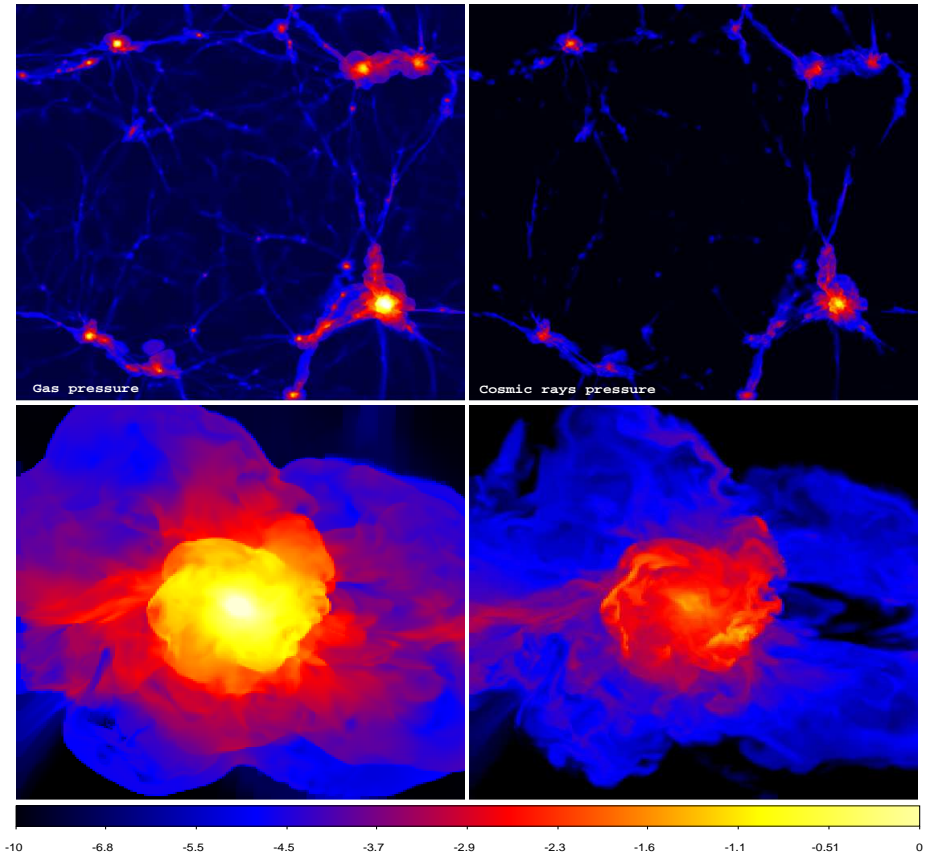}
\caption{Maps of gas and CR pressure for the simulated volume with the side of 80 Mpc (top panels) and for the AMR resimulation of a galaxy clusters (lower panels, side of 10 Mpc). The color coding is in log[arb.units]).}
\label{fig:fig2}
\end{figure}

We present here the results of non-radiative cosmological runs for a
concordance" $\Lambda$CDM cosmology, with parameters
$\Omega_{\rm dm}=0.226$, $\Omega_{\rm b}=0.044$ and $\Omega_{\rm \Lambda}=0.73$; the normalization of the primordial index of density fluctuations was set 
to $\sigma_{8}=0.8$. 
In our runs, we used an acceleration efficiency of \citet{kj07}; the injection of CR energy is modelled since the start of reionization in the simulated
volume (our tests however suggest that the initial epoch for the injection of CR is not an important parameters, provided that $z_{\rm cr}>>1$). In figure \ref{fig:fig2} (top panels), we report maps of the spatial distribution
of gas and CR pressure for the simulated volume of
80 Mpc at the uniform resolution of 156 kpc/h at z=0 ($512^{3}$ cells). Despite the overall morphological similarity between thermal gas and CR structures, the CR energy is much more concentrated around large-scale structures than gas energy; this is explained because CRs are injected only
in the downstream region of shocks, and because their adiabatic compression
during the infall onto collapsed structures is smaller than that of gas, due
to their softer equation of state.  The average pressure ratio between CRs and gas within the simulated volume has a maximum of $P_{\rm cr}/P_{\rm g} \sim 0.2-0.3$ at the cosmic gas critical density, and it declines smoothly towards the centre of massive galaxy clusters, approaching $P_{\rm cr}/P_{\rm g} \sim 0.05-0.1$ there (see right panel in Fig.\ref{fig:fig1}). Re-simulations of individual clusters employing adaptive mesh refinement ( with maximum resolution of $\sim 25$ kpc/h) provide an even smaller pressure ratio
for the core of clusters, $P_{\rm cr}/P_{\rm g} \sim 0.01-0.03$ (Fig.\ref{fig:fig2} and right panel of Fig.\ref{fig:fig1}). The dynamical role of CR energy is negligible for most of the simulated volume;  however, the final distribution of $\gamma_{\rm eff}$ as a function of radius from the center of structures is not uniform, and decreases towards
the outer regions, making the accretion regions of clusters more compressible. This systematically produces a slightly more relaxed matter distribution within clusters, compared to standard runs without CR injection, and this has a small but significant ($\sim 5-10$ per cent) impact on the measured X-ray emission profiles and  Sunyaev-Zeldovich effect (see Vazza et al. submitted). 

\section{Conclusions}
We presented a numerical development in the grid AMR code ENZO 1.5, which
allows us to treat at run-time  the injection of CR energy at 
shocks, their spatial advection and their dynamical (pressure) feedback on the evolution
of the baryonic gas. This approach provides an important and complementary approach
to similar techniques applied to Smoothed Particles Hydrodynamics
methods (e.g. \citealt{pf06}). Simulations employing fixed resolution or
adaptive mesh refinement have been run, leading to the detection
of small but non-negligible effects respect to standard large-scale structures simulated without CR physics. The application of 
these techniques to the problem of large-scale non-thermal emission
(as for example in radio relics) is underway.

\acknowledgments F.V. and M.B. acknowledge support
from the grant FOR1254 from the Deutsche Forschungsgemeinschaft. 
F.V. acknowledges the usage of computational resources under the CINECA-INAF 2008-2010 agreement, and at the Supercomputing Centre in J\"{u}lich, Germany.

\bibliography{vazza_f}

\begin{thebibliography}{}
\expandafter\ifx\csname natexlab\endcsname\relax\def\natexlab#1{#1}\fi
\expandafter\ifx\csname url\endcsname\relax
  \def\url#1{\texttt{#1}}\fi
\expandafter\ifx\csname urlprefix\endcsname\relax\def\urlprefix{URL }\fi
\providecommand{\eprint}[2][]{\url{#2}}

\bibitem[{{Bell}(1978)}]{bell78}
{Bell}, A.~R. 1978, \mnras, 182, 147

\bibitem[{{Br{\"u}ggen} et~al.(2011){Br{\"u}ggen}, {Bykov}, {Ryu}, \&
  {R{\"o}ttgering}}]{br11}
{Br{\"u}ggen}, M., {Bykov}, A., {Ryu}, D., \& {R{\"o}ttgering}, H. 2011, \ssr,
  71. \eprint{1107.5223}

\bibitem[{{Collins} et~al.(2010){Collins}, {Xu}, {Norman}, {Li}, \&
  {Li}}]{co11}
{Collins}, D.~C., {Xu}, H., {Norman}, M.~L., {Li}, H., \& {Li}, S. 2010, \apjs,
  186, 308. \eprint{0902.2594}

\bibitem[{{Ferrari} et~al.(2008){Ferrari}, {Govoni}, {Schindler}, {Bykov}, \&
  {Rephaeli}}]{fe08}
{Ferrari}, C., {Govoni}, F., {Schindler}, S., {Bykov}, A.~M., \& {Rephaeli}, Y.
  2008, \ssr, 134, 93. \eprint{0801.0985}

\bibitem[{{Kang} \& {Jones}(2007)}]{kj07}
{Kang}, H., \& {Jones}, T.~W. 2007, Astroparticle Physics, 28, 232.
  \eprint{0705.3274}

\bibitem[{{Miniati}(2007)}]{mi07}
{Miniati}, F. 2007, Journal of Computational Physics, 227, 776.
  \eprint{arXiv:astro-ph/0611499}

\bibitem[{{Norman} et~al.(2007){Norman}, {Bryan}, {Harkness}, \&
  {Bordner}}]{no07}
{Norman}, M.~L., {Bryan}, G.~L., {Harkness}, R., \& {Bordner}, J.~a. 2007,
  ArXiv e-prints, 705. \eprint{0705.1556}

\bibitem[{{Pfrommer} et~al.(2006){Pfrommer}, {Springel}, {En{\ss}lin}, \&
  {Jubelgas}}]{pf06}
{Pfrommer}, C., {Springel}, V., {En{\ss}lin}, T.~A., \& {Jubelgas}, M. 2006,
  \mnras, 367, 113. \eprint{arXiv:astro-ph/0603483}

\bibitem[{{Ryu} et~al.(2003){Ryu}, {Kang}, {Hallman}, \& {Jones}}]{ry03}
{Ryu}, D., {Kang}, H., {Hallman}, E., \& {Jones}, T.~W. 2003, \apj, 593, 599.
  \eprint{arXiv:astro-ph/0305164}

\bibitem[{{Skillman} et~al.(2008){Skillman}, {O'Shea}, {Hallman}, {Burns}, \&
  {Norman}}]{sk08}
{Skillman}, S.~W., {O'Shea}, B.~W., {Hallman}, E.~J., {Burns}, J.~O., \&
  {Norman}, M.~L. 2008, \apj, 689, 1063. \eprint{0806.1522}

\bibitem[{{Vazza} et~al.(2009){Vazza}, {Brunetti}, \& {Gheller}}]{vbg09}
{Vazza}, F., {Brunetti}, G., \& {Gheller}, C. 2009, \mnras, 395, 1333.
  \eprint{0808.0609}

\end{thebibliography}
\end{document}